\documentclass[aps,prl,superscriptaddress,twocolumn,amsmath,amssymb,floatfix]{revtex4-1}
\usepackage[utf8]{inputenc}
\usepackage[T1]{fontenc}
\usepackage{braket}
\usepackage{graphicx}
\usepackage{hyperref,xcolor}
\hypersetup{colorlinks = true, citecolor = blue, linkcolor = red}

\usepackage{amssymb}
\usepackage{amsmath}
\usepackage{natbib}
\usepackage{tikz}
\usepackage{placeins}
\usepackage{subcaption}
\usepackage{algorithm}
\def\tr{\mbox{tr}}

\newcommand{\xikj}[2]{\ensuremath{\xi^{(#1)}_{#2}}}
\newcommand{\trans}[3]{\ensuremath{\braket{#1|#2|#3}}}
\newcommand{\avg}[1]{\ensuremath{\langle\hspace{-1mm}\langle{#1}\rangle\hspace{-1mm}\rangle}}

\def\BraVert{\egroup\,\mid\,\bgroup}

\definecolor{RoyalBlue}{rgb}{0., 0.14, 0.6}

\begin{document}
	\title{Algorithmic Primitives for Quantum-Assisted Quantum Control}
	\author{Guru-Vamsi Policharla}
	\affiliation{Department of Physics, Indian Institute of Technology-Bombay, Powai, Mumbai 400076, India.}
	\author{Sai Vinjanampathy}
	\email{sai@phy.iitb.ac.in}
	\affiliation{Department of Physics, Indian Institute of Technology-Bombay, Powai, Mumbai 400076, India.}
	\affiliation{Centre for Quantum Technologies, National University of Singapore, 3 Science Drive 2, Singapore 117543, Singapore}

	\date{\today}
	\begin{abstract}
	We discuss two primitive algorithms to evaluate overlaps and transition matrix time series, which are used to construct a variety of quantum-assisted quantum control algorithms implementable on NISQ devices. Unlike previous approaches, our method bypasses tomographically complete measurements and instead relies solely on single qubit measurements. We analyse circuit complexity of composed algorithms and sources of noise arising from Trotterization and measurement errors.
	
	\end{abstract}
	\maketitle
	\paragraph{Introduction.---}
	Quantum control is central to the design of quantum technologies \cite{glaser2015training}. The control problem usually involves optimising a cost function that incorporates conditions such as distance to a target state, bandwidth and fluence restrictions. The problem is then solved by employing gradient search methods \cite{tannor1992control, khaneja2005optimal,gross1992optimal,szakacs1994locking,zhu1998rapid}, non-gradient search methods \cite{zahedinejad2016designing,spiteri2018quantum,schafer2020differentiable, niu2019universal, bukov2018reinforcement,caneva2011chopped,rach2015dcrab, palittapongarnpim2017learning,pechen2006teaching,grivopoulos2003lyapunov}  or hybrid algorithms \cite{goerz2015hybrid}. Such solutions typically involve several iterative updates to the controls before convergence to a local optima. Since these methods simulate quantum evolution on a classical computer repeatedly, often the time complexity for many-body quantum control protocols is dominated by the corresponding complexity of the evolution. 

	Several techniques to partially overcome the computational intractability have been developed. For instance, mean-field and
	cumulant expansion theories \cite{PhysRevA.100.022337,jager2014optimal} have been
	used in the place of the full evolution. 
	However, the approximations are either limited or uncontrolled. Another recent approach \cite{doria2011optimal} has been to 
	use matrix product state approaches embedded inside control algorithms.
	Though this simulates a larger set of quantum states on a classical 
	computer, such techniques cannot efficiently compute generic quantum 
	evolution.

	Given the advances in noisy intermediate-scale quantum computing (NISQ)
	\cite{boixo2018characterizing,pednault2017breaking} platforms, a 
	potential solution is to simulate quantum evolution on NISQ devices and
	extract control pulses via measurements. This problem has been 
	considered before, though the solutions have been limited. For instance, hybrid GRAPE algorithm was proposed \cite{li2017hybrid,dive2018situ} and implemented \cite{lu2017enhancing} for optimal control. The evolution of states is done on a quantum simulator and the cost function and its gradients are estimated by projecting the
	final density matrix into specific quantum states. This technique is 
	limited by the fact that the number of projectors needed to span an 
	arbitrary density matrix basis is exponential in the number 
	of qubits. To mitigate this, previous authors typically restrict the target states to sparse matrices in the measurement basis, which in turn restricts the 
	possible control solutions one can obtain.

	In this manuscript, we propose the first control solution for state 
	optimisation that is applicable to dense target states by adapting existing techniques from quantum computation and expanding them to compose 
two \textit{algorithmic primitives}. They can be implemented on a universal quantum computer, provided the many-body target states are available as an offline resource. This demand for an 
	offline resource is justified since our method is akin to compilation 
	of quantum circuits, where an optimal gate decomposition of a given 
	unitary in terms of a universal gate set is sought. Our method is 
	related to these quantum gate compilation techniques \cite{yu2013five,vartiainen2004efficient,shende2006synthesis} in seeking optima 
	but are significantly different since the underlying unitary is also 
	being optimized for a fixed target state.
	
\begin{center}
        \begin{figure}[h!]
    		\centering
    		\includegraphics[width=0.48\textwidth]{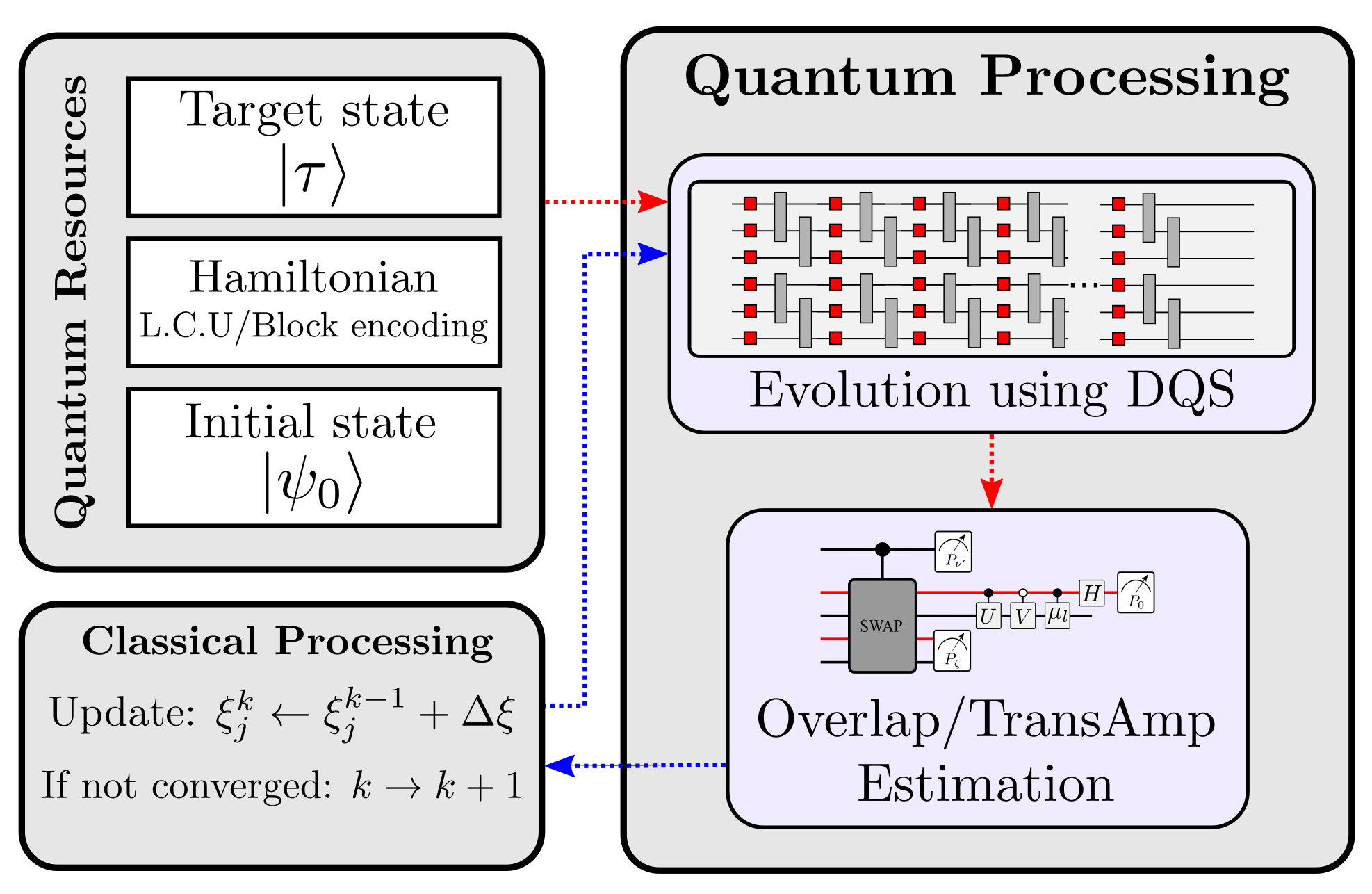}
    		\caption{A high level schematic of our quantum-assisted quantum control algorithms where scalars are extracted using OEA and TAEA given target and Hamiltonian resources.}
    		\label{fig:high_schem}
    	\end{figure}
    \end{center}

	The main insight of our work is that most optimal control techniques
	such as Krotov \cite{tannor1992control}, GRAPE 
	\cite{khaneja2005optimal}, CRAB 
	\cite{caneva2011chopped,rach2015dcrab} and machine learning methods 
	\cite{zahedinejad2016designing} only require certain scalars in 
	their update step (as opposed to a description of the entire 
	state). We propose schemes to extract these scalars, which are either in the form of the overlap of two states $\braket{\psi(t) | \chi(t)}$ or the transition element between two states in a 
	fixed Hermitian operator of the form
	$\bra{\chi(t)}\mu\ket{\psi(t)}$ using a digital quantum simulator (DQS) and qubit measurements. A schematic illustration of our quantum-assisted 
	quantum control algorithm can be found in 
	Fig.(\ref{fig:high_schem}).
    
\begin{center}
    \begin{figure}[htp!]
        \centering
        \includegraphics[width=0.48\textwidth]{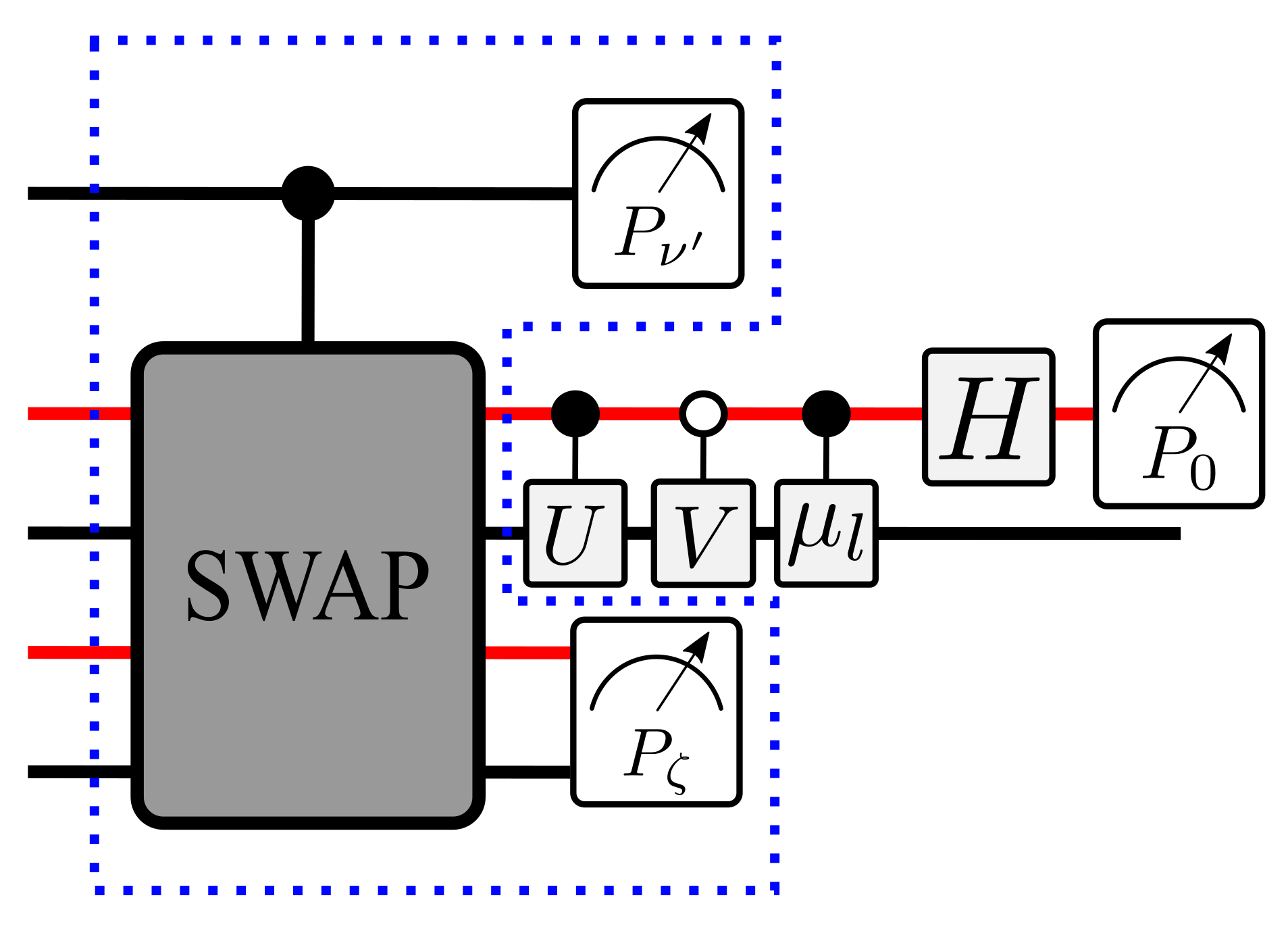}
        \caption{A circuit for estimating transition amplitudes of the form $\trans{\tau}{V^\dagger\mu U}{\psi_0}$ using the LCU technique. Using $D$ conditional unitaries $\mu_l$, the transition amplitude can be estimated. The circuit inside the dotted box is an implementation of $\Lambda_{\mathrm{sup}}$ for superposing states where $\ket{\nu'} = |\braket{\psi|\zeta}|\ket{0}+|\braket{\chi|\zeta|}|\ket{1}$.}
        \label{fig:circ-super}
    \end{figure}
    \end{center}

	\paragraph{Overlap Estimation Algorithm (OEA).---}
	Consider two time-dependent many-body quantum states $\ket{\psi(t)}=U(t)\ket{\psi_0}$ and $\ket{\chi(t)}=V(t)\ket{\chi_0}$ generated by the evolution operators which are Trotterized implementations of a time-continuous control sequence discussed below. Given these two (generic) states, we wish to estimate the overlap $\langle{\chi(t)}\vert{\psi(t)}\rangle$.
	It is well known that if we were given the bipartite state $\ket{x}:=\big(\ket{0}\ket{\psi(t)}+\ket{1}\ket{\chi(t)}\big)/\sqrt{2}$, the real part of the overlap $\langle{\chi(t)}\vert{\psi(t)}\rangle$ can be evaluated by measuring the detection probability of $\ket{1}$ on the first (ancillary) qubit of the state $H\otimes \mathbb{I}\ket{x}$, where $H$ denotes the Hadamard gate. Likewise, the imaginary part of the overlap can be obtained by starting with $\ket{x'}:=(\ket{0}\ket{\psi(t)}-i\ket{1}\ket{\chi(t)})/\sqrt{2}$.
We next outline a method to generate such superpositions.
To create superpositions between unknown states, we first extend the results of Oszmaniec et al. \cite{oszmaniec2016creating}, who proposed a probabilistic CP map $\Lambda_{\mathrm{sup}}$ to superpose any two given states $\ket{\psi}$ and $\ket{\chi}$ which have a non-zero overlap with a given reference state $\ket{\zeta}$. Concretely, their algorithm takes as input $\ket{\nu} = \alpha\ket{0}+\beta\ket{1}, \ket{\psi}$ and $\ket{\chi}$ to generate states of the form
	\begin{align}
	    \alpha\frac{\braket{\psi|\zeta}}{{|\braket{\psi|\zeta}}|}\ket{\psi} + \beta\frac{\braket{\chi|\zeta}}{|\braket{\chi|\zeta}|}\ket{\chi},
	\end{align}
    using well known post-selection methods \cite{laflamme}, as incorporated in Fig.(\ref{fig:circ-super}).
    If the overlaps are known, then the additional phase factor can be removed by modifying $\alpha$ and $\beta$ appropriately. Our 
	extension generates arbitrary superpositions of states $\ket{\psi(t)} = U(t)\ket{\psi}$ and $\ket{\chi(t)} = V(t)\ket{\chi}$ as follows: If a reference state $\ket{\zeta}$ and overlaps $\braket{\chi|\zeta} \neq 0$ and $\braket{\psi|\zeta} \neq 0$ are known,
	$\Lambda_{\mathrm{sup}}$ can be used to create a probabilistic superposition of the form
	$\alpha\ket{0}\ket{\psi} + \beta\ket{1}\ket{\chi}$ by choosing a reference state $\ket{+}\ket{\zeta}$ and inputs $\ket{0}\ket{\psi}$ and $\ket{1}\ket{\chi}$. Then by applying conditional unitaries $\ket{0}\bra{0}\otimes U(t) + \ket{1}\bra{1}\otimes V(t)$, we obtain states of the form
	\begin{equation}
	    \ket{\Psi} = \alpha \ket{0} U(t)\ket{\psi} + \beta \ket{1} V(t)\ket{\chi}.
	\end{equation}
We note that superpositions of $U(t)\ket{\psi}$ and $ V(t)\ket{\phi}$ can be created by applying a Hadamard on the first qubit and measuring it in the standard basis. If $\alpha = |\braket{\psi|\zeta}|/(\sqrt{2} \braket{\psi|\zeta})$ and $\beta = |\braket{\chi|\zeta}|/(\sqrt{2} \braket{\chi|\zeta})$, the output is the desired state $\ket{x}=(\ket{0}\ket{\psi(t)}+\ket{1}\ket{\chi(t)})/\sqrt{2}$. 
Choices of the generic quantum states specify various control 
algorithms discussed below. For instance, the choice of 
$\ket{\psi}\equiv\ket{\psi_0}$ to be the initial quantum state and
$\ket{\chi}\equiv\ket{\tau}$ to be the target state respectively will specify 
the Krotov algorithm discussed below. This method can also be used to construct several other examples of gradient-based and gradient-free control algorithms. In all such control algorithms, we 
choose the reference state $\ket{\zeta}$ to be a sparse quantum state with non-zero overlap with the initial and target state. For instance,  $\ket{\zeta}$ can be chosen to be a sparse initial state if it has non-zero overlap with the target state. We note that this choice of the reference state is not unique and can be chosen according to experimental convenience. This reduces the problem of superposition of two unitarily rotated states to the problem of applying arbitrary control unitary on an unknown state.  

Following Zhou \textit{et. al.,} \cite{zhou2011adding} we can add control to arbitrary
unitaries $U$ and $V$ by padding them with controlled-$X_a$ or \textit{internal-SWAP} 
gates. The $X_a$ gate has been experimentally demonstrated for photonic systems 
\cite{li2017hybrid} and requires that each qubit state be the lower energy manifold of 
an otherwise controllable multi-level system. This is naturally also the case for 
transmon qubits \cite{koch2007charge,gambetta_houck,solano,trimon} and several 
other physical implementations of universal quantum computers. For a four level system,
the $X_a$ gate is defined by the transformation rules $\ket{0} \leftrightarrow \ket{2},  \ket{3} \leftrightarrow \ket{1}$. For larger 
Hilbert spaces it can be defined via a SWAP gate that acts internally on subspaces of 
the same Hilbert space, distinguishing it from the usual SWAP gate that acts on tensor 
product spaces. This doubles the internal Hilbert space dimensionality of the physical 
qudit with the number of additional gates needed to add control scaling linearly with the number of 
subsystems.

\paragraph{Transition Amplitude Estimation Algorithm (TAEA).---} The second timeseries we require for the control algorithms is the estimation of transition matrix elements of the form $\bra{\chi(t)}\mu\ket{\psi(t)}$. If the Hamiltonian $\mu$ can be expressed as a linear combination of known unitaries \cite{childs2012hamiltonian} $\mu = \sum_{l=1}^{D}c_l\mu_l$, then by applying a conditional unitary $\mu_l$ on $\ket{\Psi}$ we can estimate the overlap $\trans{\chi(t)}{\mu_l}{\psi(t)}$ and hence $\trans{\chi(t)}{\mu}{\psi(t)}$. This is presented in Fig.~(\ref{fig:circ-super}) in terms of a circuit description.

Alternatively, if there exists a block encoding \cite{low2019hamiltonian,gilyen2019quantum} $B \in \mathbb{C}^k\otimes \mathcal{H}$ of $\mu \in \mathcal{H}$, then by applying a conditional version of this unitary on $({\ket{0}\ket{0_k}\ket{\psi(t)} + \ket{1}\ket{0_k}\ket{\chi(t)}})/{\sqrt{2}}$, we can estimate $\bra{\psi(t)}\bra{0_k}B\ket{0_k}\ket{\chi(t)}$ and hence $\bra{\psi(t)}\mu\ket{\chi(t)}$ (See appendix A for a formal algorithmic description of these primitives).

	\paragraph{Quantum-Assisted Gradient Algorithms.---} Let us apply these algorithmic primitives to construct quantum-assisted quantum control algorithms. Consider quantum control algorithms that operate with fixed terminal time {$(T)$} such as Krotov's algorithm \cite{tannor1992control}. Without loss of generality we work with Hamiltonians of the form $H(t) = H_0 + \xi(t)\mu$ where $H_{0}(\mu)$ is the bare(control) Hamiltonian. The cost function in the Krotov algorithm is given by
	\begin{align}\label{eq: Krotov}
	J = \bra{\psi(T)}Q\ket{\psi(T)}-\alpha \int_{0}^{T}dt\xi^2(t),
	\end{align}
	where $Q=\ket{\tau}\bra{\tau}$ is the projector onto the target state $\ket{\tau}$. Here the integral represents the fluence of the control field $\xi(t)$ and $\alpha$ is a Lagrange multiplier that modulates the relative importance of the second term. The first order variation of the cost function produces three equations of motion
	\begin{align}
	\ket{\dot{\psi}^{(k)}(t)}=&-i{H^{(k)}}\ket{\psi^{(k)}(t)}\label{eq:psi_evol}, \\
	\ket{\dot{\chi}^{(k)}(t)}=&-i{H^{(k)}}\ket{\chi^{(k)}(t)}\label{eq:chi_evol},\\
	\Delta \xi^{(k)}(t)=&-\frac{1}{\alpha}\mathrm{Im}\bra{\chi^{(k-1)}(t)}\mu\ket{\psi^{(k)}(t)}\label{eq:xi_evol},
	\end{align}
	which are solved iteratively subject to the boundary conditions $\ket{\psi^{(k)}(0)}=\ket{\psi_0}$ and $\ket{\chi^{(k)}(T)}=Q\ket{\psi^{(k)}(T)}$ until convergence is reached. Here $H^{(k)}(t)=H_0 + \xi^{(k)}(t)\mu$ is the time-dependent Hamiltonian with the 
	control field $\xi^{k}(t)$, state $\ket{\psi^{(k)}(t)}$ and co-state $\ket{\chi^{(k)}(t)}$ corresponding to the 
	$k^{th}$ iteration of the algorithm. Each 
	iteration of the solution proceeds by evolving the state and co-state 
	equations alongside computing the overlap integral. Note that the state 
	equation Eq.~(\ref{eq:psi_evol}) is specified by an initial condition whereas the co-state equation Eq.~(\ref{eq:chi_evol}) is
	specified by a terminal condition. This implies that while the state equation 
	is evolved forward in time, the co-state equation is evolved backward in time. 
	
	By using OEA and TAEA we can implement a quantum-assisted Krotov algorithm that circumvents the exponential complexity of simulating state evolution. Throughout the algorithm we only use superpositions of the form $\ket{y} =\alpha\ket{0}\ket{\psi(0)}+\beta\ket{
	1}\ket{\tau}$ which can be prepared as described earlier if the overlaps with respect to some reference state are known. Evolution of states is done according to Eq.~(\ref{eq:psi_evol}) and Eq.(\ref{eq:chi_evol}) via digital quantum simulation which can be done efficiently even for many body systems. 
	{Note that since 
	$\ket{\chi(t)}=V(t)\ket{\chi(T)}=\braket{\tau| \psi(T)} 
	V(t)\ket{\tau}$, we need to estimate the scalar overlap $\langle\tau\vert 
	\psi(T)\rangle$ which can be done using OEA given $\ket{y}$.}
	To compute the control field updates 
	$\Delta\xi^{(k)}(t)$ we employ TAEA. 
	Other popular algorithms such as GRAPE can also be implemented in a quantum assisted fashion (see appendix B).
	
    \paragraph{Quantum-Assisted Gradient-free Algorithms.---} Besides gradient algorithms, OEA and TAEA can be used to also implement non-gradient quantum control algorithms. As an example, we consider the Chopped Random Basis (CRAB) Optimisation \cite{caneva2011chopped} which works by performing a search on a truncated basis. The candidate control field is written as
	$\xi(t) = \sum_{i=1}^N c_i\xi_i(t)$ where $\xi_i$'s are typically trigonometric functions. The coefficients $c_i$ are optimised using standard non-gradient techniques such as Nelder-Mead algorithm \cite{nelder1965simplex} to obtain the optimal pulse sequences. Since the Nelder-Mead technique uses the function to be optimised in a black-box fashion the only quantity that needs to be extracted is the fidelity which is done using a SWAP test \cite{buhrman2001quantum}. Likewise, the dCRAB technique \cite{rach2015dcrab} and machine learning based control algorithms such as SuSSADE \cite{zahedinejad2016designing} can be assisted by a quantum simulator since the only relevant quantity needed for optimization is fidelity.
    
	\paragraph{Resource Counting---}
	Our algorithmic primitives use $m$ repeated single-qubit measurements to estimate scalar values, which we group into a single \emph{experiment} and begin the complexity analysis with the number of experiments needed for several example algorithms. We then separately estimate the relationship between Trotter error, number of measurement repetitions and the error threshold per experiment. The superposition part of either primitive requires measurements in a basis containing the reference state $\ket{\zeta}$ which can be done easily if a unitary transformation can be implemented such that it takes some standard basis state to $\ket{\zeta}$.
	The OEA requires two experiments, one each for estimating real and imaginary parts. The number of experiments for TAEA depends on the approach used. While block encoding only needs two experiments, 
	LCU requires $2D$ experiments one each for the real and imaginary parts of $D$ different terms in the LCU decomposition of the operator $\mu$. 
	The number of times each quantum-assisted control algorithm invokes the algorithmic primitives can be inferred simply. For each update, one TAEA experiment and two OEA experiments are needed. Quantum-assisted GRAPE requires two fidelity experiments in each update step.

   We now relate the error thresholds to the number of measurements $m$ and pulse updates $n$ by focusing on the quantum-assisted Krotov algorithm, the extension of this analysis to 
    other algorithms being straightforward. We consider two sources of error, the
    first is due to Trotterization which results in an approximate version of the 
    unitary corresponding to time evolution being implemented and the second is 
    due to the finite precision arising from a finite number of measurements. 
    Since the Krotov algorithm has the property of monotonic convergence, the 
    quantum-assisted Krotov algorithm enjoys this property as well up until the 
    point where the incremental change in the cost function $\Delta J^{(k)}= J^{(k)} - 
    J^{(k-1)}$ in an ideal implementation is comparable to the fluctuation in 
    $\Delta J^{(k)}$ due to the two sources of errors mentioned above. Poulin et al. 
    \cite{poulin2011quantum} upper bound the error of a Hamiltonian written as 
    the sum of $L$ terms $H(t) = \sum_{i = 1}^{L} H_i(t)$ with bounded norm. Assuming each term acts on at most $\kappa$ qubits, the 
    total error in approximating the exact time evolution is upper bounded by 
    \begin{equation}
        ||U - U^{\mathrm{TS}}|| = \varepsilon_{TS} \leq \frac{c_{max}^2 T^2}{2n^2}
    \end{equation}
    where $n=T/\Delta t$, $T$ is the total time of evolution, $\Delta t$ is the time interval over which Hamiltonian fluctuation is negligible, $||\cdot ||$ denotes Hamiltonian norm, $c_{max} = \mathrm{max}_t\mathrm{max}_i||H_i(t)||$, $U = \mathcal{T}\exp\{-i/\hbar 
    \int_0^TH(t)dt\}$ is the time ordered integral and $U^{\mathrm{TS}}$ is the 
    Trotter-Suzuki implementation of $U$. The number of gates required to 
    implement the Trotterized gates is given by 
    $d_{sk}G(\varepsilon_{TS},L,T)\log(G(\varepsilon_{TS},L,T)/\varepsilon_{TS})^{c_{sk}}$ where 
    $d_{sk},c_{sk}$ are Solovay-Kitaev constants and 
    $G(\varepsilon_{TS},L,T)=c_{max}^2T^2L^3/\varepsilon_{TS}$. Thus choosing $\varepsilon_{TS}$ fixes the total number of gates in the Trotter decomposition. We note that though we present upper limits to the number of gates, many-body localization was recently shown to stabilize local observables \cite{heyl2019quantum} for time-independent DQS with relatively small number of gate operations while retaining desired levels of accuracy, a result that directly applies to our protocols.
    
    A second source of error we consider is due to the finite precision arising from a finite number of measurements. The number of runs of TAEA and OEA relates to the fluctuations in the pulse update $\Delta \xikj{k}{j}$ and can be used to upperbound it using $\varepsilon_M$ (see appendix C). Using this we are able to upperbound the fluctuations in $\Delta J^{(k)}$ due to the two sources of error. As long as the stochastically averaged change is larger than than the maximum fluctuation we are guaranteed monotonicity of the quantum-assisted Krotov algorithm. This condition can be written as
\begin{equation}
	        \avg{\Delta J^{(k)}} \gg 4T\varepsilon{||\mu||} + 4\frac{T}{\alpha}[\xi_{max}\varepsilon] + 4\varepsilon_{TS}\label{eq:deltajk}
\end{equation}
where $T$ is total time and $\xi_{max}$ is the maximum value of the control field, $\varepsilon = \varepsilon_M + 3\varepsilon_{TS}||\mu||/\alpha$.
Hence by setting a threshold on the stochastically averaged cost function $\avg{\Delta J^{(k)}}$ below which we stop optimization, we can compute the required number of gates and measurements in terms of $\varepsilon_{TS}$ and $\varepsilon_M$ respectively. We note that for bounded Hamiltonians Eq.~(\ref{eq:deltajk}) suggests a reasonable scaling of number of gates and measurements needed to perform quantum-assisted quantum control.
\paragraph{Conclusions.---}
We present the algorithmic primitives OEA and TAEA to implement arbitrary quantum-assisted quantum control algorithms. Our method substantially improves existing hybrid quantum control algorithms by incorporating dense target states and complex control algorithms.  Furthermore, the underlying algorithmic primitives only rely on qubit measurements and require the implementation of specific controlled unitaries, that can be accomplished by well known methods. We emphasize that the circuit complexity of specific algorithms considered scales favourably with error thresholds. We applied our method to quantum-assisted versions of gradient and non-gradient algorithms. Several recent experiments have demonstrated control of relatively large number of qubits on differing platforms, from gradient algorithm on twelve NMR qubits \cite{lu2017enhancing} to a controlled state preparation of twenty transmon qubits \cite{Song574} suggesting that our protocols are practical to implement with existing technology. Furthermore, when two separate platforms can both simulate the same Hamiltonians $H_0$ and $\mu$, our method can be combined with analog/digital quantum simulation experiments to design optimal control pulses in new systems. For example, proposals to generate all-optical higher-dimensional tensor network states \cite{Walmsley_Tensor,Ish_Tensor} can be used to generate complex target states for our scheme. 

Our techniques add to the literature on variational quantum algorithms \cite{khatri2019quantum,PhysRevX.7.021027,PhysRevA.100.022327} and generalize the optimization program to complex overlaps which bring control theory objectives and variational quantum algorithms closer \cite{magann2020pulses}. Furthermore, our algorithmic primitives can also be used to optimize non-gradient objectives and can be used for the important tasks such as the design of modular quantum computers \cite{devitt2009architectural,arenz2018controlling,lee2020upper} and controlling reactions in quantum chemistry \cite{peruzzo2014variational}.

\paragraph{Acknowledgements.---} SV acknowledges support from the DST-SERB Early Career Research Award (ECR/2018/000957) and DST-QUEST grant number DST/ICPS/QuST/Theme-4/2019. SV thanks Kishor Bharti, Adolfo del Campo, Ish Dhand, Carlos Perez-Delgado and Dario Poletti for insightful discussions.
	
%

	\color{black}
	\clearpage
	\appendix
	\onecolumngrid
\section{A. Description of Algorithmic Primitives}
Here we describe in detail the algorithms for estimating overlaps (Algorithm \ref{alg:OEA}) and transition amplitudes using LCU (Algorithm \ref{alg:transampLCU}) and Block Encoding (Algorithm \ref{alg:transampBE}).
    \begin{algorithm}[H]
	\caption{OEA}\label{alg:OEA}
	\begin{enumerate}
	    \item Given states $\ket{x} = (\ket{0}\ket{\psi(t)} + \ket{1}\ket{\chi(t)})/\sqrt{2}$ and $\ket{x'} = \ket{0}\ket{\psi(t)} - i\ket{1}\ket{\chi(t)}/\sqrt{2}$.
	    \item Apply $H\otimes \mathbb{I}$ on $\ket{x}$ and $\ket{x'}$.
	    \item Measure the ancilla qubits of $\ket{x}$ and $\ket{x'}$ in the standard basis and estimate the probability of obtaining $\ket{1}$ by performing $m$ repetitions. Let these probabilities be $P_x$ and $P_{x'}$ respectively.
	    \item The overlap $\braket{\psi(t)|\chi(t)}$ can be computed as $$\mathrm{Re}(\braket{\psi(t)|\chi(t)}) = 2P_x - 1,$$ $$\mathrm{Im}(\braket{\psi(t)|\chi(t)}) = 2P_{x'} - 1.$$
	\end{enumerate}
    \end{algorithm}
    
    \renewcommand{\thealgorithm}{2.\arabic{algorithm}}
    \setcounter{algorithm}{0}
	\begin{algorithm}[H]
	\caption{TAEA via LCU}\label{alg:transampLCU}
	\begin{enumerate}
	    \item Given a linear combination of unitaries $\mu = \sum_{j=1}^D c_j U_j$.
	    \item Estimate the overlap for each $U_j$ $$t_r^{(j)},t_i^{(j)} \gets \textsc{Overlap}(\ket{\chi(t)},U_j\ket{\psi(t)}).$$
		\item Combine these values to obtain 
		$$\mathrm{Re}(\trans{\chi(t)}{\mu}{\psi(t)}) \gets \sum_{j = 1}^{k} c_j t_r^{(j)},$$
		$$\mathrm{Im}(\trans{\chi(t)}{\mu}{\psi(t)})  \gets \sum_{j = 1}^{k} c_j t_i^{(j)}.$$
	\end{enumerate}
    \end{algorithm}
    \begin{algorithm}[H]
	\caption{TAEA via Block Encoding}\label{alg:transampBE}
	\begin{enumerate}
	    \item Given an $(\alpha,k,\epsilon)$-block encoding $B$ of $\mu$ defined according to \cite{gilyen2019quantum}.
	    \item Create the superposition $$({\ket{0}\ket{0_k}\ket{\psi(t)} + \ket{1}\ket{0_k}\ket{\chi(t)}})/{\sqrt{2}},$$
	    and apply a conditional unitary $B$.
	    \item The transition amplitude can be computed as $\trans{\chi(t)}{\mu}{\psi(t)} = \alpha\big( \textsc{Overlap}(\ket{0_k}\ket{\chi(t)},B\ket{0_k} \ket{\psi(t)})\big)$. 
	\end{enumerate}
    \end{algorithm}
\section{B. Quantum-Assisted GRAPE Algorithm}
As another application of our algorithmic primitives, we discuss the implementation of the quantum-assisted GRAPE algorithm. GRAPE discretizes the time varying pulse into constant pulses over small time intervals of size $\Delta t$. Each piece-wise constant step is updated iteratively in a manner that monotonically increases the fidelity between the target state and evolved state. The unitary corresponding to the pulse sequence in the interval $j\Delta t \leq t < (j+1) \Delta t$ is $U_j = \exp\{-i\Delta t\big(H_0 + \sum_{k=1}^L u_k(j\Delta t) H_k \big)\}$. If we start with a state $\rho_0$, the final evolved state is $\rho(T) = U_L\dots U_0 \rho_0 U_0^\dagger \dots U_L^\dagger$. Over each iteration, the update to $u_k(j\Delta t)$ is given by $\Delta u_k(j\Delta t) \propto {\partial \phi_0}/{\partial u_k(j\Delta t)}$, where $\phi_0$ is the quantity to minimize (infidelity say). The derivative can be written as \cite{khaneja2005optimal}
	$$\frac{\partial \phi_0}{\partial u_k(j)} = - \tr({\chi_j^\dagger i \Delta t [H_k,\rho_j]})$$
	where
	\begin{align*}
	\rho_j &= U_j \dots U_0 \rho_0 U_0^\dagger \dots U_j^\dagger  \\
	\chi_j &= U_{j+1}^\dagger \dots U_n^\dagger \rho_{tar} U_n \dots U_{j+1}
	\end{align*}
	and $\rho_{tar}$ is the target state.
	GRAPE suffers from the same limitation as the Krotov method, which is that exponentially large resources are required to compute the evolution for large system sizes. However, this can be done in polynomial time even for many body systems by using DQS and extracting the gradients. The derivative can be computed as the difference between two SWAP tests \cite{buhrman2001quantum}
	
	\begin{align}
	\tr(\chi_j^\dagger i \Delta t [(H_0 + H_k), \rho_j]) - \tr(\chi_j^\dagger i \Delta t [H_0,\rho_j])\approx
	 -\tr(\chi_j^\dagger e^{-i \Delta t (H_0 + H_k)}\rho_je^{i \Delta t (H_0 + H_k)}) +\tr(\chi_j^\dagger e^{-i \Delta t (H_0)}\rho_je^{i \Delta t (H_0)}),\label{eq:GRAPE}
	\end{align}
	where terms of order higher than $\Delta t$ have been ignored for the approximation.
	\section{C. Uniform Convergence Threshold for Noisy Quantum-Assisted Krotov Algorithm}
	We now show that the quantum-assisted Krotov algorithm is monotonically convergent until the updates are of the same magnitude as the errors. Recall the cost function \begin{equation}J^{(k)} = |\braket{\tau|U^{(k)}|\psi(0)}|^2 - \alpha\Delta t\sum_{j=1}^n (\xikj{k}{j})^2 \end{equation}
	where $k$ is the iteration number and $\xikj{k}{j}$ is the value of the pulse between $(j-1)\Delta t$ and $j\Delta t$. The time evolution is implemented via DQS as a result of which an approximate unitary $W$ is implemented in place of the actual unitary. We can write the incremental change of the cost function in our algorithm as
		\begin{align}
		    \Delta J^{(k)} &= |\braket{\tau|W^{(k)}|\psi(0)}|^2 - |\braket{\tau|W^{(k-1)}|\psi(0)}|^2
		- \alpha\Delta t\sum_{j=1}^n [(\xikj{k}{j})^2 - (\xikj{k-1}{j})^2]
		\end{align}
	The pulse update is estimated by running the TAEA and OEA and then computing:
	\begin{equation}
	    \Delta \xikj{k}{j} = \frac{-1}{\alpha}\mathrm{Im}\big(\braket{\psi^{(k-1)}(T)|\tau} \braket{\tau|W_{j,n}^{(k-1)}\mu W_{1,j}^{(k)}|\psi(0)}\big)
	\end{equation}
	where $W_{n_1,n_2}^{(k)}$ is the Trotterized implementation of $U_{n_1,n_2}^{(k)} = \prod_{m=n_1}^{n_2}\exp(-i \Delta tH^{(k)}(m\Delta t))$.
	Let $\braket{\psi^{(k-1)}(T)|\tau} = a_1 + i b_1$, where $a_1$ and 
	$b_1$ are estimated using single qubits measurements in OE and hence are Guassian random variables with a variance of $\sigma_b^2/m$, where 
	$\sigma_b^2$ is the variance of the Bernoulli random variable 
	corresponding to the outcome of a projective measurement on the qubit 
	and $m$ is the number of measurements or samples. Similarly, if 
	$\braket{\tau|W_{j,n}^{(k-1)}\mu W_{1,j}^{(k)}|\psi(0)} = a_2 + ib_2$, 
	then the variance of $a_2,b_2$ depends on the technique used for TAE. 
	If the LCU technique is used it is $(\sum_{l=1}^D c_l^2)\sigma_b^2/m = c^2\sigma_b^2/m$. Here we have ignored 
	second order terms that originate from errors introduced in the current iteration. Then
	$$\Delta \xikj{k}{j} = -\frac{1}{\alpha}(a_1b_2 + a_2b_1)$$
	By using the properties of variances ($\mathrm{Var}(XY) = \mathrm{Var}(X)\mathrm{Var}(Y) + \mathrm{Var}(X)\mathrm{Eva}(Y)^2 + \mathrm{Var}(Y)\mathrm{Eva}(X)^2$) it can be seen that the variance of $\Delta \xikj{k}{j}$ is
	\begin{align}
		\mathrm{Var}(\Delta \xikj{k}{j}) &= \frac{\sigma_b^2}{\alpha^2m}\big(c^2\avg{a_1}^2 + \avg{a_2}^2 + c^2\avg{b_1}^2 + \avg{b_2}^2 + \frac{2c^2\sigma_b^2}{m}\big) \nonumber\\
		&\leq \frac{1}{4\alpha^2m}\big(c^2 + ||\mu||^2\big)
	\end{align}
	where we used $|\braket{\tau|W_{j,n}^{(k-1)}|\mu|W_{1,j}^{(k)}|\psi(0)}|^2 \leq ||\mu||^2$, $|\braket{\psi^{(k-1)}(T)|\tau}|^2 \leq 1$, $\sigma_b^2 \leq 1/4$ and ignored ${c^2}/{2m}$ for the inequality.

We are yet to account for the error in the updates due to Trotterization. Since $||U_{n_1,n_2}^{(k)} - W^{(k)}_{n_1,n_2}|| = ||\Delta U_{n_1,n_2}^{(k)}|| \leq \varepsilon_{TS}$, 
\begin{align}\braket{\tau|U_{j,n}^{(k-1)}\mu U_{1,j}^{(k)}|\psi(0)} &= \braket{\tau|W_{j,n}^{(k-1)}\mu W_{1,j}^{(k)}|\psi(0)} + \braket{\tau|\Delta U_{j,n}^{(k-1)} \mu W_{1,j}^{(k)}|\psi(0)} +\\
&\quad \braket{\tau|W_{j,n}^{(k-1)}\mu \Delta U_{1,j}^{(k)}|\psi(0)} + \braket{\tau|\Delta U_{j,n}^{(k-1)}\mu \Delta U_{1,j}^{(k)}|\psi(0)}.
\end{align}
We know that $|\braket{a|\mu \Delta U|b}|\leq \varepsilon_{TS}||\mu||$, then

\begin{align}
    &\quad |\mathrm{Im}\big(\braket{\psi(0)|U^{(k-1)\dagger}|\tau} \braket{\tau|U_{j,n}^{(k-1)}\mu U_{1,j}^{(k)}|\psi(0)}\big) - \mathrm{Im}\big(\braket{\psi(0)|W^{(k-1)\dagger}|\tau} \braket{\tau|W_{j,n}^{(k-1)}\mu W_{1,j}^{(k)}|\psi(0)}\big)|\\
    &\leq |\braket{\psi(0)|U^{(k-1)\dagger}|\tau}\braket{\tau|U_{j,n}^{(k-1)}\mu U_{1,j}^{(k)}|\psi(0)}- \braket{\psi(0)|W^{(k-1)\dagger}|\tau}\braket{\tau|W_{j,n}^{(k-1)}\mu W_{1,j}^{(k)}|\psi(0)}|\\
    &\leq 3\varepsilon_{TS}||\mu|| \quad \text{(Ignoring higher order terms)}
\end{align}
Thus, an error $\delta_j^k$ such that $ |\delta_j^k| \leq 3\varepsilon_{TS}|\mu||/\alpha$ is introduced in the updates due to Trotterization.

We can write $\Delta \xikj{k}{j} = \avg{\Delta \xikj{k}{j}} + \epsilon^k_j + \delta_j^k$ where $\epsilon^k_j$ is a random variable with the same distribution as $\Delta \xikj{k}{j}$ but mean shifted to 0 and $\avg{\circ}$ denotes the stochastic average over noise realisations (Trotterization and measurement errors). 
Then from Chebyshev's inequality 
\begin{align}
    \mathrm{Pr}[|\epsilon_j^k|\geq\varepsilon_M] \leq \frac{\mathrm{Var}(\epsilon_j^k)}{N\varepsilon_M^2} \leq \frac{1}{4\alpha^2m N \varepsilon_M^2}\big(c^2 + ||\mu||^2 \big)\label{eq:chebyshev}
\end{align}
where $N$ is the number of samples of $\epsilon_j^k$ used for averaging. 
We have thus bounded both sources of errors. 

These small fluctuations result in a slightly different unitary (before Trotterization) being chosen that can be described as follows
	\begin{align}
		U^{(k)} &= \prod_{j=1}^{n} e^{i\Delta t[H_0 + (\avg{\xikj{k}{j}} + \epsilon_j^k + \delta_j^k))\mu]} \\
		&\approx \prod_{j=1}^{n} e^{i\Delta t[H_0 + (\avg{\xikj{k}{j}})\mu]}(\mathbb{I} +  i\mu \Delta t(\epsilon_j^k+ \delta_j^k))) \\
		&= \avg{U^{(k)}} + \sum_{j = 1}^n i(\epsilon_j^{k}+ \delta_j^k)\Delta t \mu^{j,k}
	\end{align}
	where we have ignored all terms of order larger than $\Delta t (\epsilon_j^k + \delta_j^k)$ and $$\mu^{j,k} = \big(\prod_{j'=1}^{j-1}e^{i\Delta t[H_0 + (\avg{\xikj{k}{j'}})\mu]}\big) \mu \big(\prod_{j'=j+1}^{n}e^{i\Delta t[H_0 + (\avg{\xikj{k}{j'}})\mu]}\big)$$

	The incremental change in cost function can then be written as
		\begin{align}
			\Delta J^{(k)} 	=&|\braket{\tau|W^{(k)}|\psi(0)}|^2 - |\braket{\tau|W^{(k-1)}|\psi(0)}|^2
		- \alpha\Delta t\sum_{j=1}^n [(\xikj{k}{j})^2 - (\xikj{k-1}{j})^2]\\ 
			=& |\braket{\tau|\avg{U^{(k)}}|\psi(0)} + \braket{\tau|{{W^{(k)}} - U^{(k)}}|\psi(0)} + \sum_{j=1}^ni(\epsilon_j^k + \delta_j^k) \Delta t \trans{\psi(0)}{\mu^{j,k}}{\tau}|^2\nonumber\\
			&-|\braket{\tau|\avg{U^{(k-1)}}|\psi(0)} + \braket{\tau|{{W^{(k-1)}} - U^{(k-1)}}|\psi(0)} + \sum_{j=1}^ni(\epsilon_j^{k-1} + \delta_j^{k-1}) \Delta t \trans{\psi(0)}{\mu^{j,k-1}}{\tau}|^2 \nonumber\\
			&-\frac{\Delta t}{\alpha}\sum_{j=1}^{n}[(\avg{\xikj{j}{k}} + \epsilon_j^k + \delta_j^k)^2 - (\avg{\xikj{j}{k-1}} + \epsilon_j^{k-1}  + \delta_j^{k-1})^2]\\
			\geq& |\braket{\tau|\avg{U^{(k)}}|\psi(0)}|^2 + |\braket{\tau|{{W^{(k)}} - U^{(k)}}|\psi(0)}|^2 - |\braket{\tau|\avg{U^{(k-1)}}|\psi(0)}|^2 - |\braket{\tau|{{W^{(k-1)}} - U^{(k-1)}}|\psi(0)}|^2 \nonumber\\
			&+|\sum_{j=1}^n(\epsilon_j^k +\delta_j^k) \Delta t \trans{\psi(0)}{\mu^{j,k}}{\tau}|^2 - |\sum_{j=1}^n(\epsilon_j^{k-1}  + \delta_j^{k-1}) \Delta t \trans{\psi(0)}{\mu^{j,k-1}}{\tau}|^2 \nonumber\\
			& -2|\braket{\tau|\avg{U^{(k)}}|\psi(0)}||\braket{\tau|{{W^{(k)}} - U^{(k)}}|\psi(0)}| - 2|\braket{\tau|\avg{U^{(k-1)}}|\psi(0)}||\braket{\tau|{{W^{(k-1)}} - U^{(k-1)}}|\psi(0)}|\nonumber\\
			&-2|\braket{\tau|\avg{U^{(k)}}|\psi(0)}||\sum_{j=1}^n(\epsilon_j^k +\delta_j^k) \Delta t \trans{\psi(0)}{\mu^{j,k}}{\tau}|\nonumber\\
			&-2|\braket{\tau|\avg{U^{(k-1)}}|\psi(0)}||\sum_{j=1}^n(\epsilon_j^{k-1}+\delta_j^{k-1}) \Delta t \trans{\psi(0)}{\mu^{j,k-1}}{\tau}| \nonumber\\
			&-2|\braket{\tau|{{W^{(k)}} - U^{(k)}}|\psi(0)}||\sum_{j=1}^n(\epsilon_j^k +\delta_j^k)\Delta t \trans{\psi(0)}{\mu^{j,k}}{\tau}|\nonumber\\
			&-2|\braket{\tau|{{W^{(k-1)}} - U^{(k-1)}}|\psi(0)}||\sum_{j=1}^n(\epsilon_j^{k-1}+\delta_j^{k-1})\Delta t \trans{\psi(0)}{\mu^{j,k-1}}{\tau}| \nonumber\\
			&-\frac{\Delta t}{\alpha}\sum_{j=1}^{n}[(\avg{\xikj{j}{k}})^2 - (\avg{\xikj{j}{k-1}})^2 + (\epsilon_j^k + \delta_j^k)^2 - (\epsilon_j^{k-1} + \delta_j^{k-1})^2 + 2\avg{\xikj{j}{k}}(\epsilon_j^k + \delta_j^k) - 2\avg{\xikj{j}{k-1}}(\epsilon_j^{k-1} + \delta_j^{k-1})]
		\end{align}
where we used $|a|^2 + |b|^2 + |c|^2 - 2|a||b| - 2|b||c| - 2|c||a|\leq |a+b+c|^2 \leq (|a|+|b|+|c|)^2$ for the inequality. In \cite{tannor1992control}, it was shown that the (error-free) Krotov algorithm is monotonically convergent \cite{suri2018speeding}, namely
\begin{align}
\avg{\Delta J^{(k)}} = |\braket{\tau|\avg{U^{(k)}}|\psi(0)}|^2 - |\braket{\tau|\avg{U^{(k-1)}}|\psi(0)}|^2 -\frac{\Delta t}{\alpha}\sum_{j=1}^{n}[(\avg{\xikj{j}{k}})^2 - (\avg{\xikj{j}{k-1}})^2] > 0.
\end{align}
However, if the fluctuations in $\Delta J^{(k)}$ due to measurement and Trotter errors, overwhelm $\avg{\Delta J^{(k)}}$, it is no longer possible to reliably achieve better results. Concretely, if
\begin{align}
	        \avg{\Delta J^{(k)}} \gg 4T(\varepsilon_M + 3\varepsilon_{TS}||\mu||/\alpha){||\mu||} + 4\frac{T}{\alpha}[\xi_{max}(\varepsilon_M + 3\varepsilon_{TS}||\mu||/\alpha)] + 4\varepsilon_{TS}\label{eq:monotonicity}
\end{align}
where $\xi_{max} = \mathrm{max}_j(\xikj{k}{j},\xikj{k-1}{k})$ is the maximum amplitude of the pulse (can be upperbounded by experimental constraints), then with very high probability the fluctuations are much smaller than the actual update and monotonicity is guaranteed. Eq.~(\ref{eq:monotonicity}) follows from the fact that the variance is upperbounded by $\varepsilon_M$ with high probability and $||{W} - U|| \leq \epsilon_{TS}$. We have also dropped terms of higher order from the expression. Writing $\varepsilon = \varepsilon_M + 3\varepsilon_{TS}||\mu||/\alpha$, we get Eq.(8) in the main text.

In practice, one would first choose how ``close'' they wish to get to a target state by fixing a value $\Delta J_{min}$ beyond which they stop the quantum control optimization. Once this is fixed, $\varepsilon_M$ and $\varepsilon_{TS}$ are chosen subject to experimental constraints using which the number of measurements/gates needed can be calculated using Eq.~(\ref{eq:chebyshev}) and the results on Trotterization from \cite{poulin2011quantum} (see section on robustness in main text).

\end{document}